\documentstyle[prb,aps]{revtex}
\begin{document}
\twocolumn[\hsize\textwidth\columnwidth\hsize\csname@twocolumnfalse%
\endcsname
\title{Coulomb Ordering in Anderson-Localized Electron Systems.}
\author{A.~A.~Slutskin$^1$, M.~Pepper$^2$, and H.~A.~Kovtun$^1$.}
\address{$^1$Institute for Low Temperature Physics and
Engineering, 47 Lenin Ave., Kharkov, Ukraine.\protect \\
$^2$Cavendish Laboratory, University of Cambridge, Madingley Road,
Cambridge CB3 0HE, UK}
\maketitle
\date{\today}
\draft
\begin{abstract}
We consider an electron system under conditions of strong Anderson
localization, taking into account interelectron long-range Coulomb
repulsion. We have established that with the electron density going to
zero the Coulomb interaction brings the arrangement of the Anderson
localized electrons closer and closer to an ideal (Wigner) crystal
lattice, provided the temperature is sufficiently low and the dimension
of the system is $> 1$. The ordering occurs despite the fact that a
random spread of the energy levels of the localized one-electron
states, exceeding the mean Coulomb energy per electron, renders it
impossible the electrons to be self-localized due to their mutual
Coulomb repulsion This differs principally the Coulomb ordered Anderson
localized electron system (COALES) from Wigner crystal, Wigner glass,
and any other ordered electron or hole system that results from the
Coulomb self-localization of electrons/holes.  The residual disorder
inherent to COALES is found to bring about a multi-valley ground-state
degeneration akin to that in spin glass. With the electron density
increasing, COALES is revealed to turn into Wigner glass or a glassy
state of a Fermi-glass type depending on the width of the random spread
of the electron levels.

\end{abstract}
 ]

\paragraph{\bf Introduction.} \label{Intro} As was shown by Wigner long
ago\cite{Wigner}, slow decrease in the Coulomb electron-electron
interaction potential $v(r) = e^2/\kappa r$ ($e$ is the free electron
charge, $r$ is a distance between the interacting electrons, $\kappa$
is the permittivity) with an increase in the interelectron distance
inevitably causes the Coulomb energy of free electron gas to exceed its
kinetic energy at sufficiently low electron densities with the
resulting transition of the gas into an electron crystal (Wigner
crystal). In the wake of Wigner's prediction a natural question was
raised whether long-range (weakly screened) Coulomb forces can lead to
ordering of charge carrier ensembles in conductors. However, a strong
evidence of "Wigner-crystal-in-crystal" existence has not been found
yet. This suggests that the Wigner crystallization, at least in the
convential conductors, is very difficult (if at all possible) to
observe in pure form. Therefore, seeking mechanisms of charge carrier
Coulomb ordering that are beyond the above Wigner's scenario is of
great interest.

At present much attention is being given to an electron/hole Coulomb
self-localization in lattice systems with so small overlap intergal $t$
that tunnelling of charge carriers is supressed by their mutual Coulomb
interaction. The self-localization brings about electron/hole ordering
in two cases: i/the host lattice is regular (generally an
incommensurate electron/hole structure is formed\cite{Hubbard,SSK}),
ii/the host lattice is disordered, but the mean separation of its sites
is much less than that of the charge carriers $\bar r$ (the so-called
Wigner glass is formed, whose space structure, though disordered, is
close to a Wigner crystal lattice (WCL)\cite{note} in a
sense\cite{Shkl1}).  Ordered self-localized charge carrier lattice
systems (OSLCCLS) of both types are obviously not the same as Wigner
crystal. In addition, unlike the Wigner crystallization, the higher is
the electron/hole density the weaker are dynamic effects caused by the
charge carrier tunnelling.

The present paper aims to draw attention to a new type of Coulomb
ordering that occurs under conditions of a strong Anderson
localization. Specifically, we deal with Anderson localized electron
systems wherein the external random field localizing the electrons is a
random set of potential wells that are sufficiently deep for the
localization length $r_L$ to be $\lesssim$ the mean wells separation,
$r_0 \sim n_0^{-1/d}$ ($n_0$ is the density of the wells, $d$ is the
dimension of the system). We show further that at sufficiently low
electron densities, $n_e$, and sufficiently low temperatures, $T$,  the
mutual Coulomb repulsion of the localized  electrons  inevitably forces
them to be arranged close to the WCL sites, provided $d >$~1.  Such
Coulomb ordered Anderson localized electron system (COALES) differs
qualitatively from both Wigner crystal and OSLCCLS for the following
reason. An electron/hole self-localization underlying formation of
Wigner crystal or OSLCCLS occurs only if the {\em Coulomb energy of
mutual charge carrier repulsion is the prevailing one}. In the electron
systems undergoing strong Anderson localization this is not the case, as
the width of {\em random spread of the electron levels in the wells},
$\Delta$ (the disorder energy) is much more than the mean Coulomb
energy per electron $\varepsilon_c \sim v(\bar r)$. Under these
conditions a Coulomb self-localization is impossible since a shift of
an electron over a small distance $\sim r_0 \ll \bar r$ changes the
electron energy much more than $\varepsilon_c$. Below we describe a new
mechanism by which Coulomb ordering occurs despite the inequality
$\varepsilon_c \ll \Delta$.

The smallness of $\varepsilon_c$ as compared with $\Delta$ is believed
to be a good reason to describe an influence of Coulomb
electron-electron interaction on the systems with strong Anderson
localization in terms of Fermi glass,\cite{Mott} a random system of the
electrons occupying all wells with the energies $\leq$ the Fermi energy
$\varepsilon_F$.  Up to now, it is the Fermi-glass approximation
modified with regard to existence of the so-called Coulomb gap
(Efros-Shklovskii gap)\cite{Shkl2} at the Fermi energy has been the
basic approach to the problem. COALES existence at low $n_e$ is beyond
the scope of the Fermi-glass concept. This indicates that the
convential line of reasoning should be revised.

COALES posseses a peculiar ``duality'':  similar to Wigner
crystal, it arises as $n_e \rightarrow 0$, while its space structure
similar to that of Wigner glass. As will be seen from the following,
such ``duality'' not only differs COALES from both Wigner crystal and
Wigner glass but imparts it a number of new features that make COALES
an appealing subject of investigation. We also show that with an
increase in $n_e$ COALES turns into Fermi glass or OSLCCLS depending on
a value of the parameter $\gamma = \Delta/v(r_0)$, governing an
interplay between the random spread of the electron levels and the
Coulomb interelectron repulsion. The residual disorder inherent to
COALES is described here in terms of {\em dipole glass}, a random
system of interacting dipoles that issue from the WCL sites. The dipole
representation of the disorder proves to be helpful for revealing a
multi-valley degeneration of the COALES ground state reminiscent of
that in the spin glasses.

\paragraph{\bf The Basic Assumptions.}\label{Basic} Owing to the
inequality $r_L \lesssim r_0$ the radius-vectors $\vec R$ of the
centers of the random-potential wells can be considered as quantum
numbers of the one-electron localized states, $|\vec R>$. Their
energies $\varepsilon (\vec R)$ are random values, the structure of the
$\vec R$ set is assumed to be arbitrary, in particular, it can be
regular.  Without essential loss of generality, the density $\nu$ of
the number of $\varepsilon (\vec R)$ with a given value $\varepsilon$
can be presented in the form

\begin{equation}
\label{nu}
\nu = n_0f(\varepsilon)/\Delta\,,
\end{equation}

\noindent where $f(\varepsilon)$ is $\sim 1$ within the interval
$[\varepsilon_{\text {min}},\,\varepsilon_{\text {min}} + \Delta]$ and
equals zero outside it; $\varepsilon_{\text {min}}$ is the least of the
$\varepsilon (\vec R)$. Further we put $\varepsilon_{\text {min}} = 0$,
$f(0) = 1$.

Here it is assumed that $\bar r \gg r_0$. In particular, this allows to
neglect perturbations produced in the eigenstates of the system by the
electron-electron interaction (with accuracy to additions $\sim
(r_0/\bar r)^2 \exp(-r_0/r_L) \ll 1$). In this approximation, the
eigenstates can be identified with

\begin{equation}
\label{eigenstates}
|{\cal R}> = |\vec R_1>|\vec R_2>\ldots |\vec R_N>\,.
\end{equation}

\noindent where ${\cal R} = \vec R_1,\ldots,\vec
R_N$ denotes a set of the wells occupied by the
electrons, $N$ is the number of the electrons. The complete set of the
eigenstates comprises all possible ${\cal R}$
with different $\vec R_1\ldots \vec R_N$. With the same accuracy, the
eigenvalues $E({\cal R})$ corresponding to the states $|{\cal R}>$ take
the form

\begin{mathletters}
\label{Ham}
\begin{equation}
\label{H}
E({\cal R}) = E_c({\cal R}) + E_w({\cal R})\,,
\end{equation}

\noindent where $E_c$ and $E_w$ are the energy of the mutual electron
repulsion and the energy of the non-interacting Anderson localized
electrons in the wells, respectively:

\begin{equation}
\label{E_c_exc}
E_c = \frac12\sum_{i,k=1\atop i\neq k}^N
v(|\vec R_i - \vec R_k|), \quad \quad  E_w =
\sum_{i=1}^{N} \varepsilon(\vec R_i)\,;
\end{equation}
\end{mathletters}

\noindent  We aim to find the structure of the electron configuration
${\cal R}_g$ that minimizes $E({\cal R})$.

\paragraph{\bf Electron Ordering and Fermi-Glass Instability.}
\label{ordering} First let us discuss how mutual electron repulsion
affects the ground state of the Fermi-glass in the limit $n_e
\rightarrow\, 0$.  As follows from Eq.(\ref{nu}), $n_e$ is {\em linear}
in $\varepsilon_F$ at $\varepsilon_F\ll \Delta$: $$n_e = n_0
\varepsilon_F/\Delta\,,$$

\noindent  On the
other hand, $\varepsilon_c$ is proportional to $n_e^{1/d}$. Hence, the
ratio $\varepsilon_F/\varepsilon_c \propto n_e^{\frac{d-1}{d}}$ tends
to {\em zero} (for $d>1$) with a decrease in $n_e$. What this means is
that Fermi glass with sufficiently low $n_e$ is {\em unstable} with
respect to mutual Coulomb repulsion of electrons.

If the electrons were free to move and $\varepsilon(\vec R) = 0$, the
ground-state configuration ${\cal R}_g$ would be a WCL. Since the
electrons of Fermi glass are {\em randomly} arranged, the Coulomb
energy per electron of the Fermi-glass exceeds that of the WCL,
$\varepsilon_0 \sim \varepsilon_c$, significantly (i.e. by a value
$\sim \varepsilon_c$). This suggests that for sufficiently low $n_e$
the configuration ${\cal R}_g$ falls into a class of $\cal R$ that meet
the folowing conditions:  i/ for each WCL site there is an electron
located in a small neighborhood of the site, ii/the upper bound
$\varepsilon_b$ of the electron energies in the wells $\varepsilon(\vec
R_1),\ldots,\varepsilon(\vec R_N)$ satisfies the inequalities

\begin{equation}
\label{inequalities}
\varepsilon_F \ll \varepsilon_b \ll \varepsilon_c\,,
\end{equation}

\noindent The energy per electron $\varepsilon_{\cal R} = E({\cal
R})/N$ of such $\cal R$ is near $\varepsilon_0$. It cannot be
less than $\varepsilon_0$ as the mean electron energy in the wells
$\varepsilon_w = E_w/N\geq \varepsilon_F$.

Any configuration of the above class belongs to the set of points $\vec
R$ for which $\varepsilon(\vec R) \leq \varepsilon_b$. The density of
these points equals $n_0(\varepsilon_b/\Delta)$, and their mean
separation

\begin{equation}
\label{delta_r}
\rho(\varepsilon_b) \sim  r_0 (\Delta/\varepsilon_b)^{1/d}\,,
\end{equation}

\noindent is much less than the WCL spacing $a_0 \sim \bar r$ owing to
the first of the inequalities (\ref{inequalities}). Therefore, for each
WCL site $\vec m$ there are inevitably several $\vec R$ of the set such
that $|\vec R - \vec m| \sim \rho$.  Populating these ``proximate''
states $|\vec R>$ (one electron per site) yields just the
configurations of the class we are interested in. For such a
configuration the energy $\varepsilon_{\cal R}$ is the sum

\begin{mathletters}
\label{energy}
\begin{equation}
\label{energy1}
\varepsilon_{\cal R} = \varepsilon_0 + \delta \varepsilon\,,
\end{equation}

\noindent where $\delta \varepsilon$ is expected to be a small
correction to $\varepsilon_0$. It consists of two terms:

\begin{equation}
\label{delta}
\delta \varepsilon =
a\,\left(\rho/\bar r\right)^2 \varepsilon_0 +
b\,(r_0/\rho)^d\Delta
\end{equation}
\end{mathletters}

\noindent The first term is the deformation energy produced by
electron displacements over distances $\sim \rho$ from the WCL
sites. The second term is $\varepsilon_w$ expressed in terms of $\rho$
in view of Eq.(\ref{delta_r}) and the fact that $\varepsilon_w \sim
\varepsilon_b$.  Factors $a = a({\cal R}),b = b({\cal R})$ depend on
details of ${\cal R}$, but they are both $\sim 1$ for any of the
considered configurations.

The ground-state correction, $\delta \varepsilon_g$, to $\varepsilon_0$
is the least of the $\delta \varepsilon$ values. Putting $a,b = 1$ in
the expression (\ref{delta}) and finding its minimum in $\rho$, we
obtain the estimate

\begin{equation}
\label{min_correction}
\delta \varepsilon_g \sim
\gamma^{\frac{2}{d+2}}
\left(r_0/\bar r \right)^{\frac{2(d-1)}{d+2}}\varepsilon_0\,,
\end{equation}

\noindent  the minimum being reached at

\begin{equation}
\label{rho_min}
\rho = \bar\rho \sim
\gamma^{\frac{1}{d+2}}
\left(r_0/\bar r \right)^{\frac{d-1}{d+2}}a_0
\sim  \left(\Delta/\delta E_c \right)^{\frac{1}{d+2}}r_0\,,
\end{equation}

\noindent where $\gamma$ is the parameter defined in item \ref{Intro};
the physical sense of the energy $\delta E_c = (r_0/\bar
r)^2\varepsilon_c$ will be explained below (item \ref{existence}). The
quantity $\bar\rho$ characterizes deviation of ${\cal R}_g$ from WCL.

Expressions (\ref{min_correction}) and (\ref{rho_min}) show that
$\delta \varepsilon_g/\varepsilon_0$ and $\bar\rho/a_0$ both go to zero
when $\bar r/r_0 \rightarrow\, \infty$. Hence, for sufficiently low
$n_e$ the ground-state electron configuration is WCL slightly perturbed
by random electron displacements from the WCL sites. This is just the
COALES mentioned in the item \ref{Intro}. The space structure of COALES
differs from that of Wigner glass\cite{Shkl1} at an important point:
the typical displacement of the electrons/holes from the WCL sites in
Wigner glass is of order of the geometrical constant of the system, the
mean host-lattice sites separation, while that in COALES, $\bar\rho$,
depends not only on the geometrical parameter $r_0$ but also on both
the disorder energy $\Delta$ and the electron density.

\paragraph{\bf Dipole Glass.} \label{Dip_gl} In order to do justice to
the COALES ground state it is necessary to describe the residual
disorder inherent to this system, extending the consideration to cover
all ${\cal R}$ that are close to ${\cal R}_g$. To this end, it is
convenient to introduce the effective Hamiltonian $H_{\text{eff}}$ by
the formula $E({\cal R}) = N\varepsilon_0 + H_{\text{eff}}$, expressing
$H_{\text{eff}}$ in terms of dipoles $\vec d_{\vec m} = \vec R - \vec
m$ as independent variables ($\vec m$ are the radius-vectors of the WCL
sites) whose moduli are $\ll a_0$. Expanding $E_c$ in powers of
$d_{\vec m}^{\alpha}$ (index $\alpha$ enumerates the components of
vectors) and restricting to quadratic in $\vec d_{\vec m}$ terms, it is
easy to obtain

\begin{equation}
\label{dipoles}
H_{\text{eff}} = \sum_{\vec m}\tilde\varepsilon(\vec
d_{\vec m})\, + \,\frac12\sum_{\vec m,\vec
m\,'}\Lambda_{\alpha\alpha'}(\vec m - \vec m\,')d^{\alpha}_{\vec
m}d^{\alpha'}_{\vec m\,'}\,,
\end{equation}

\noindent where matrix $\Lambda_{\alpha\alpha'}(\vec m) = |\vec
m|^{-3}(\delta_{\alpha\alpha'} - 3m_{\alpha}m_{\alpha\,'}/|\vec m|^2)$
determines the interaction of two dipoles issuing from sites $0\,, \vec
m$; the function $\tilde\varepsilon(\vec d_{\vec m}) = \varepsilon(\vec
m + \vec d_{\vec m}) + \bar\Lambda_{\alpha\alpha'}d^{\alpha}_{\vec
m}d^{\alpha'}_{\vec m}$, the matrix $\bar\Lambda_{\alpha\alpha'} =
-\sum_{\vec m}\Lambda_{\alpha\alpha'}(\vec m)$; here and further on
summation over repeated indexes $\alpha,\alpha'$ is implied, the sums
of $\vec m, \vec m\,'$ are taken over all WCL sites. The second term in
$\tilde\varepsilon(\vec d_{\vec m})$ is the energy of interaction of a
given dipole with WCL; the second term in Eq.(\ref{dipoles}) is the
energy of the dipole-dipole interaction.

The distinctive feature of $H_{\text{eff}}$ is that the dipoles are
variables taking on their values on a given {\em random} set. This
suggests that the considered dipole system has much in common with the
spin glass \cite{spin_glass}. The analogy becomes clear for the
simplest model in which $\vec d_{\vec m} = d_0 s_{\vec m} \vec e_{\vec
m} $, $s_{\vec m} = \pm 1$, $d_0$ is a constant, and unit vectors
$\vec e_{\vec m}$ constitute a given random set, $H_{\text{eff}}$
taking the form

\begin{equation}
\label{spins}
H_{\text{eff}} =
\sum_{\vec m \neq \vec m\,'}J_{\vec m \vec m\,'}s_{\vec m}s_{\vec
m\,'}\, + \,\sum_{\vec m} h(s_{\vec m})\,,
\end{equation}

\noindent where $s_{\vec m}$ are the independent variables, the
``exchange integral'' $J_{\vec m \vec m\,'} = d_0^2
\Lambda_{\alpha\alpha'}(\vec m - \vec m\,')e^{\alpha}_{\vec m}
e^{\alpha'}_{\vec m\,'}$ is a {\em random} matrix, and $h(s_{\vec m}) =
\tilde\varepsilon(d_0 s_{\vec m} \vec e_{\vec m})$ plays the role of an
external random field. The system with the Hamiltonian (\ref{spins}),
being a special case of spin glass \cite{spin_glass}, shares with the
spin glasses their known general property: a multi-valley ground-state
degeneration. It can be described in terms of ``${\cal
N}$-excitations'' that are $s_{\vec m}$ configurations differing from
the ground-state one by a big number ${\cal N}$ of spin flips.  The
multi-valley degeneration implies that separation of the low bound of
the ${\cal N}$-excitations energy spectrum from the ground-state energy
is $\sim$ the typical separation between the neighboring energies of
the spectrum, $\delta E_{\cal N} \propto {\cal N}/ Z$ ($Z$ is the total
number of ${\cal N}$-excitations with a given ${\cal N}$, $\ln Z
\propto {\cal N}$), and in consequence of this, is {\em exponentially
small in ${\cal N}$}.

The true dipole Hamiltonian (\ref{dipoles}) differs from the model one
(\ref{spins}) only in that each dipole $\vec d_{\vec m}$ runs through a
given finite random set containing more than two vectors. (The vectors
of the set are such that their moduli are comparable with $\bar\rho$).
Therefore, the above properties of the energy spectrum of the ${\cal
N}$-excitations (they are dipole configurations with ${\cal N}$ dipoles
other than in the ground state) hold in the general case. In other
words, the multi-valley degeneration does take place in the dipole
system, and hence, in COALES.

\paragraph{\bf The Region of COALES Existence.}\label{existence}
Expression (\ref{rho_min}) shows that with an increase in $n_e$ the
ratio $\bar\rho/a_0$ increases, while $\bar\rho$ itself decreases.
This brings about, depending on a value of the parameter $\gamma$, two
different scenarios of what happens with COALES as $n_e$ increases. For
sufficiently big $\gamma$ the ratio $\bar\rho/a_0$, increasing together
with $n_e$, reaches inevitably some critical value $\beta < 1/2$ at
which the length of the space correlations in the system is $\sim \bar
r$, and COALES turns (supposedly, by a second order transition) into a
{\em glassy state}. This occurs at $$n_e = n_{e1} \sim
\beta_c^{\frac{d(d+2)}{d-1}} \gamma^{-\frac{d}{d-1}}n_0\,.$$

\noindent At this point the ratio $\varepsilon_F/\varepsilon_c$ is
$\sim \beta^{d+2}$, i.e. it is significantly less than $1$. Therefore,
for $n_e \geq n_{e1}$ there exists a $n_e$ range over which the glassy
state cannot be adequately described in terms of the notion of Fermi
energy. A further decrease in $n_e$ leads to transformation of such
``non-Fermi''glass into conventional Fermi glass (modified by the
Coulomb electron-electron interactions) only if $\gamma \geq 1$.
Otherwise, $\varepsilon_F/\varepsilon_c \sim \gamma
(n_e/n_0)^{\frac{d-1}{d}}$ is less than $1$ for any $n_e$, and
non-Fermi glass exists for all $n_e \geq n_{e1}$.

If $\gamma$ is sufficiently small, an increase in $n_e$ reduces
$\bar\rho$ down to its least possible value, $r_0$, the ratio
$\bar\rho/a_0$ remaining $\ll 1$.  This takes place at $\Delta \sim
\delta E_c$, or in terms of $n_e$, at $n_e = n_{e2} \sim
\gamma^{\frac{d}{3}}n_0$. The parameter $\delta E_c$ is the typical
change in the Coulomb energy of the system as an electron is shifted
over distance $\sim r_0$. Therefore, for $\delta E_c > \Delta$ the
mutual electron repulsion dominates the random spread of the electron
levels, causing localization of the electrons by itself. In other
words, in the considered case the state into which COALES turns as
$n_e$ increases is nothing but the above-mentioned OSLCCLS whose host
lattice is a regular or disordered $\vec R$ set.

It follows from the aforesaid that the region of COALES existence on
the $n_e,\gamma$-plane is restricted by $\gamma$-axis and two curves,
$n_e = n_{e1}(\gamma),\,\,n_e = n_{c1}(\gamma)$ ($n_{e1}(\gamma)
\rightarrow\, 0$ as $\gamma \rightarrow\, \infty$; $n_{e2}(0) = 0$).
They intersect at point $n_e = \bar n_e \sim \beta^d n_0, \gamma =\bar
\gamma \sim \beta^3$, the value $\bar n_e$ being the maximal $n_e$ at
which COALES exists.

Heating affects COALES if $T$ exceeds $\delta
\varepsilon_g$, both terms in Eq.(\ref{delta}) ($\rho = \bar\rho$)
being $\sim T$. This gives the dipole thermal-fluctuations amplitude
$d_T \sim \bar r (\varepsilon_c/T)^{1/2}$, the fluctuating dipole
vector taking $\sim (T/\delta \varepsilon_g)^{1+d/2}$ values. As $T$
increases, $d_T$ becomes $\sim a_0$, and at some critical $T \sim
\varepsilon_c$ COALES turns into a glassy state.

\paragraph{\bf Some General Features of COALES.}\label{general}
Macroscopically, COALES manifests both electron crystal and
glassy-state features.

Due to the proximity of the COALES space structure to WCL the low bound
of the energy spectrum of one-electron excitations produced by
displacements of an electron over distances $\gg \bar r$ is separated
from the ground state energy by a big gap $\sim \varepsilon_c$. This
renders one-electron variable-range hopping over COALES impossible.
Low-temperature conduction in COALES is by multi-electron exchange
processes of a creep type (a power dependence of the COALES
conductivity $\sigma$ on $T$ is expected) or  by transfer of charged
point defects (positively charged WCL vacancies and interstitial
electrons), $\sigma$ being proportional to their concentration that in
its turn is $\propto \exp((\varepsilon_v -\varepsilon_i)/T)$
($\varepsilon_v$ and $\varepsilon_i >\varepsilon_v$ are the energies of
vacancy and interstitial-electron formation respectively;
$\varepsilon_i - \varepsilon_v < \varepsilon_c$). Thus, $\sigma$ by no
means obeys the known Mott's low \cite{Mott} for Fermi glass.

The multi-valley degeneration of the dipole glass, similarly to that in
spin glass \cite{spin_glass}, is bound to cause an infinite spectrum of
relaxation times, and, in consequence of this, a non-ergodic behavior
of COALES. This can be revealed by observation of different relaxation
processes in COALES proceeding for anomalously long times. An example
is relaxation of a non-equilibrium COALES polarization created in one
way or another. This has much in common with relaxation of a
non-equilibrium magnetic moment in spin glass.\cite{spin_glass}

As follows from the aforesaid, under moderate-disorder conditions
($\Delta \lesssim v(r_0)$) COALES can exist up to $n_e$ that are only
several times less than $n_0$. The proper materials to observe COALES
are various amorphous narrow-band conductors, superlattices, and
inversion layers wherein $n_e$ can be varied within wide limits without
affecting the disorder. Favorable conditions for two-dimensional COALES
existence are expected to be realized in semiconductor superlattices
with the so-called $\delta$-layers. Of special interest is a conductive
sheet in a system $metal$ --- $n$-type $GaAs$ --- $p$-type $GaAs$ with
charge transfer in an impurity band.\cite{Pepper} It is distinguished
by pronounced regular-type oscillations in $n_e$, which cannot be
explained in conventional one-electron terms and most likely are a
manifestation of two-dimensional COALES or OSLCCLS. We intend to
provide new evidence to confirm the suggestion and to specify the
electron structure in nearest future.

The main outlines of COALES mentioned here are to be issues of our
further detailed publications.

We gratefuly acknowledge interesting and useful discussions with
I.Lerner and I.Yurkevich. We are sincerely thankful to B.Shklovskii for
helpful remarks.

The visit of A.~Slutskin to the Cavendish Laboratory was supported by
the Royal Society.  AS would like to express his gratitude.


\begin{references}
\bibitem{note} To avoid confusion, it should be noted that here and
further on we use this term, for the sake of shortness, only to
designate the electron/hole ideal crystal lattice that would realize
the absolute minimum of the Coulomb electron/hole repulsion energy if
the electrons/holes were free to move.
\end{references}
\end{document}